\documentclass[aps,prb,10pt,twocolumn,superscriptaddress,nobibnotes]{revtex4-1}

\usepackage{lmodern}
\usepackage[T1]{fontenc}
\usepackage[utf8]{inputenc}
\usepackage{graphicx}
\usepackage{amsmath,amssymb,amsfonts}
\usepackage[unicode=true,
            colorlinks=true,
            linkcolor=blue,
            citecolor=blue,
            urlcolor=blue]{hyperref} 
\usepackage{siunitx}
\usepackage{multirow}
\usepackage{lipsum}
\usepackage[x11names]{xcolor}

\newcommand{\domark}{%
  \vbox to 0pt{
    \kern-\dp\strutbox
    \smash{\llap{\color{red!90!black}\#\kern0.5em}}
    \vss
  }%
}

\begin{document}

\title{
Far-field Excitation of Single Graphene Plasmon Cavities with Ultra-compressed Mode-volumes
	}
	
\author{Itai Epstein}
\email{itai.epstein@icfo.eu}
\affiliation{ICFO-Institut de Ciencies Fotoniques, The Barcelona Institute of Science and Technology, 08860 Castelldefels (Barcelona), Spain}

\author{David Alcaraz}
\affiliation{ICFO-Institut de Ciencies Fotoniques, The Barcelona Institute of Science and Technology, 08860 Castelldefels (Barcelona), Spain}

\author{Zhiqin Huang}
\affiliation{Department of Electrical and Computer Engineering, Duke University, Durham, NC, 27708, USA}
\affiliation{Center for Metamaterials and Integrated Plasmonics,Duke University, Durham, NC, 27708, USA}

\author{Varun-Varma Pusapati}
\affiliation{ICFO-Institut de Ciencies Fotoniques, The Barcelona Institute of Science and Technology, 08860 Castelldefels (Barcelona), Spain}

\author{Jean-Paul Hugonin}
\affiliation{Université Paris-Saclay, Institut d'Optique Graduate School, CNRS, Laboratoire Charles Fabry, 91127, Palaiseau, France}

\author{Avinash Kumar}
\affiliation{ICFO-Institut de Ciencies Fotoniques, The Barcelona Institute of Science and Technology, 08860 Castelldefels (Barcelona), Spain}

\author{Xander M. Deputy}
\affiliation{Department of Electrical and Computer Engineering, Duke University, Durham, NC, 27708, USA}
\affiliation{Center for Metamaterials and Integrated Plasmonics,Duke University, Durham, NC, 27708, USA}

\author{Tymofiy Khodkov}
\affiliation{ICFO-Institut de Ciencies Fotoniques, The Barcelona Institute of Science and Technology, 08860 Castelldefels (Barcelona), Spain}

\author{Tatiana G. Rappoport}
\affiliation{Centro de F\'{i}sica and Departamento de F\'{i}sica and QuantaLab, Universidade do Minho, P-4710-057 Braga, Portugal}

\author{Nuno M. R. Peres}
\affiliation{Centro de F\'{i}sica and Departamento de F\'{i}sica and QuantaLab, Universidade do Minho, P-4710-057 Braga, Portugal}
\affiliation{International Iberian Nanotechnology Laboratory (INL), Av. Mestre José Veiga, 4715-330 Braga, Portugal}

\author{David R. Smith}
\affiliation{Department of Electrical and Computer Engineering, Duke University, Durham, NC, 27708, USA}
\affiliation{Center for Metamaterials and Integrated Plasmonics,Duke University, Durham, NC, 27708, USA}

\author{Frank H. L. Koppens}
\email{frank.koppens@icfo.eu}
\affiliation{ICFO-Institut de Ciencies Fotoniques, The Barcelona Institute of Science and Technology, 08860 Castelldefels (Barcelona), Spain}
\affiliation{ICREA – Instituci\'{o} Catalana de Recerca i Estudis Avan\c{c}ats, Barcelona, Spain}

\begin{abstract}
\textbf{
Acoustic-graphene-plasmons (AGPs) are highly confined electromagnetic modes, carrying large momentum and low loss in the mid-infrared/Terahertz spectra. Owing to their ability to confine light to extremely small dimensions, they bear great potential for ultra-strong light-matter interactions in this long wavelength regime, where molecular fingerprints reside. However, until now AGPs have been restricted to micron-scale areas, reducing their confinement potential by several orders-of-magnitude. Here, by utilizing a new type of graphene-based magnetic-resonance, we realize single, nanometric-scale AGP cavities, reaching record-breaking mode-volume confinement factors of $\thicksim5\cdot10^{-10}$. This AGP cavity acts as a mid-infrared nanoantenna, which is efficiently excited from the far-field, and electrically tuneble over an ultra-broadband spectrum. Our approach provides a new platform for studying ultra-strong-coupling phenomena, such as chemical manipulation via vibrational-strong-coupling, and a path to efficient detectors and sensors, in this challenging spectral range.
}
\end{abstract}
\maketitle
 
Graphene plasmons (GPs) are propagating electromagnetic waves that are coupled to electron oscillations within a graphene sheet. They have attracted great scientific interest in the last few years, owing to their extraordinary properties of extreme confinement and low loss in the mid-infrared (MIR) to terahertz (THz) spectrum \cite{Wunsch2006,Hwang2007,Jablan2009,Koppens2011,Grigorenko2012,Low2014}. This extreme confinement provides a unique platform to probe a variety of optical and electronic phenomena, such as quantum non-local effects \cite{Lundeberg2017}, molecular spectroscopy \cite{Chen2017}, and biosensing \cite{Rodrigo2015}, together with enabling the access to forbidden transitions by bridging the scale between light and atoms \cite{Rivera2016}. Furthermore, it provides means to miniaturize optoelectronic devices in the long wavelength spectrum, such as GP-based electro-optical detectors \cite{Lundeberg2017a}, modulators \cite{Woessner2017}, and electrical excitation \cite{DeVega2017,Guerrero-Becerra2019}. \\

The confinement of GPs can be increased even further by placing the graphene sheet close to a metallic surface \cite{Alonso-Gonzalez2017}. Such a structure supports a highly confined asymmetric mode, which is referred to as an acoustic-graphene-plasmon (AGP), due to its linear energy versus momentum dispersion. When the graphene-metal distance is very small, AGPs are confined in-plane extensively to almost $1/300$  their equivalent free-space wavelength \cite{Lundeberg2017}, and are vertically confined to the spacing between the metal and graphene \cite{Iranzo2018}. \\

This unmatched ability of AGPs to confine light to small dimensions is pivotal for strong light-matter interactions, especially in the MIR and THz spectra, where the wavelength is inherently large and  limits the electromagnetic field confinement. The latter is eminently important at this spectral region, as it is where molecular resonances reside, and their spectral fingerprints are needed to be detected for industrial purposes, medicine, biotechnology and security\cite{Baxter2011,Haas2016}. Yet, so far AGPs have only been observed over micron-scale areas, either as THz free-propagating waves\cite{Alonso-Gonzalez2017,Lundeberg2017}, or in MIR grating couplers \cite{Iranzo2018,Lee2019}. To exploit strong and ultra-strong light-matter interactions, individual compact cavities for AGPs are required.\\

In this work, we realize single, nanometric-scale AGP cavities, with ultra-small mode volumes that are $\thicksim5\cdot10^{10}$ times smaller than the free-space wavelength volume, and which efficiently couple to far-field light. This is achieved by the generation of a graphene-plasmon-magnetic-resonance (GPMR), which enables the far-field excitation of AGP cavities, over large areas, without the need for lithographic patterning of neither the surrounding nor the graphene, and with no limitations on the light polarization. The GPMRs are formed by dispersing metallic nanocubes, with random locations and orientations, over an hBN/graphene van der Waals heterostructure. We show that each nanocube forms with the graphene a single GPMR, which is associated with a direct excitation of a highly confined AGP. Furthermore, we find that this type of coupling is highly efficient and a nanocube fill-factor of only a few percentages is sufficient for obtaining a strong optical response, which is comparable to that obtained by grating couplers with much larger fill-factors\cite{Iranzo2018,Lee2019}. Finally, we show that a single GPMR structure acts as an AGP nanoantenna, for which its scattered radiation can be electrically tuned over an ultra-broadband range. \\

\begin{figure*}[p!] 
  \centering
  \includegraphics[scale=0.375]{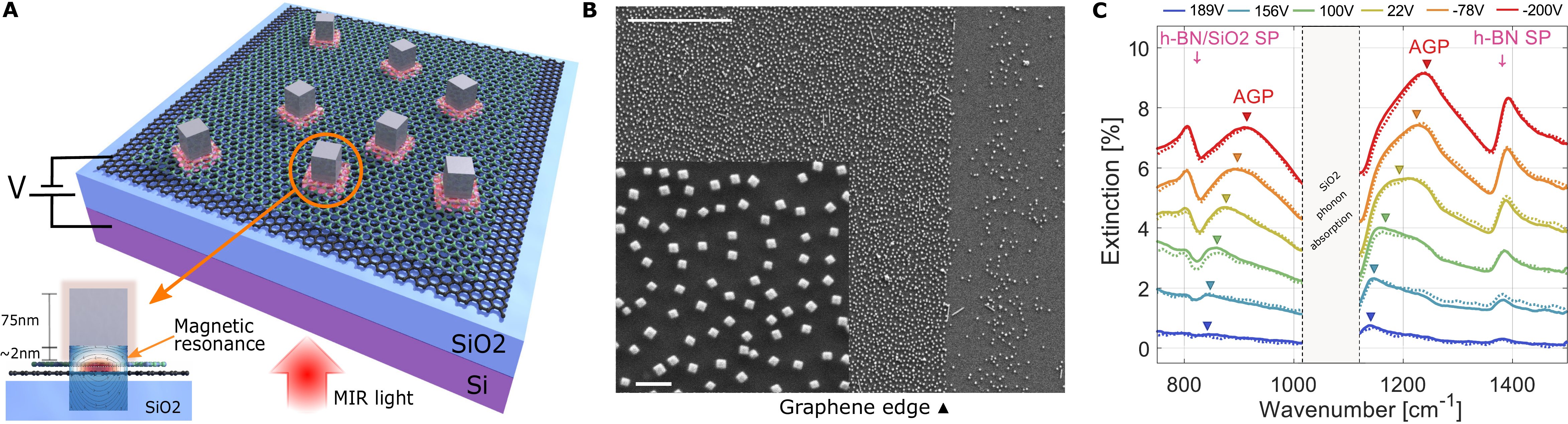} 
   \caption{\textbf{Structure and optical response of a GPMR device}. \textbf{A} Schematic of the GPMR device, which is composed of a Si/$\mathrm{SiO}_2$  substrate, graphene/h-BN heterostructre, and silver nanocubes. The inset shows the cross-section of a single GPMR structure and the generated magnetic resonance between the nanocube and the graphene. \textbf{B} SEM images of the actual device, showing the graphene edge and deposited nanocubes. The inset presents a zoomed-in image of the nanocubes (scale bars are $5$ $\mathrm{\mu m}$ and $200$ $\mathrm{nm}$, respectively). \textbf{C} A typical GPMR device extinction spectra measured in an FTIR apparatus, for different gate voltages (colors). The colored triangle marks the AGP peak and its evolution with gate voltage. The downward arrows mark the location of the hBN and $\mathrm{SiO}_2$ surface-phonons. The solid (dotted) lines correspond to the optical response for unpolarized (polarized) light. The strong abortion band of the $\mathrm{SiO}_2$  phonon is also marked. 
\label{fig:figure1}}
\end{figure*}

The scheme for a GPMR device is presented in Fig.~\ref{fig:figure1}A - nanometric silver nanocubes are randomly dispersed on top of an h-BN/graphene heterostructure that is transferred on a Si/$\mathrm{SiO}_2$  substrate \cite{Si2020}, which acts as a back-gate for electrically doping the graphene. The actual device, as imaged by a scanning-electron-microscope (SEM), is presented in Fig.~\ref{fig:figure1}B and shows the randomly scattered nanocubes over the graphene. The Fourier-transform-infrared-spectroscopy (FTIR) extinction spectra, $1-\mathrm{T(V)/T(V_{max})}$, measured from the device for different back-gate voltages (colors), is shown in Fig.~\ref{fig:figure1}C, where $\mathrm{T(V)}$ is the transmission measured at a specific gate voltage $\mathrm{V}$, and $\mathrm{T(V_{max})}$ is transmission measured for the maximal voltage (corresponding to the lowest doping due to the intrinsic doping of CVD graphene).  Two resonances can be observed in the gate-dependent extinction (marked with triangles), on both sides of the $\mathrm{SiO}_2$ phonon absorption band (marked box). These resonances are visibly shifting to higher frequencies with increasing graphene doping, which is consistent with the well-known behavior of plasmons in graphene \cite{Iranzo2018,Yan2013,Dai2015}. In addition, the supported surface-phonons in the $\mathrm{SiO}_2$ and hBN (marked with arrows) lead to the typical surface-phonon-graphene-plasmon hybridization in this type of structures \cite{Brar2014,Iranzo2018,Yan2013,Dai2015}. \\


The effect of the polarization is also shown in Fig.~\ref{fig:figure1}C. The spectra measured for both polarized (solid curves) and unpolarized (dotted curves) illumination are nearly identical. Taking into account the fact that AGPs are transverse-magnetic (TM) modes, this lack of preference on polarization in the response verifies the random nature of the nanocubes, and implies a very weak interaction between neighboring nanocubes. \\

\begin{figure*}[p!] 
  \centering
  \includegraphics[scale=0.52]{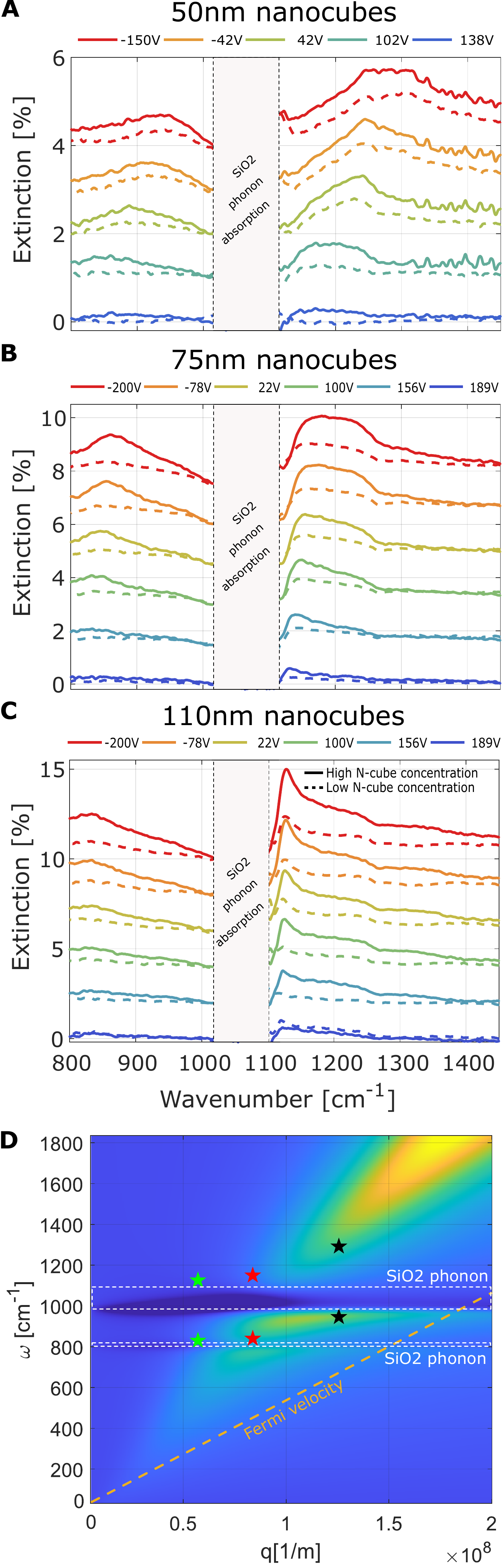} 
   \caption{\textbf{Optical response for different nanocube sizes and concentration}. Gate-dependent extinction for nanocube sizes of \textbf{A} $50$ $\mathrm{nm}$, \textbf{B} $75$ $\mathrm{nm}$ and \textbf{C} $110$ $\mathrm{nm}$, showing the change in spectral response with nanocubes dimension. The solid and dashed curves correspond to higher and lower nanocubes concentrations, respectively, and show that the change in the nanocube concentration corresponds to a change in the amplitude of the response, without affecting its spectrum. The data for different gate voltages has been shifted for clarity. For simplicity, no hBN capping layer has been used in these samples. \textbf{D} Calculated dispersion relation for a Ag/	2nm dielectric spacer/graphene/$\mathrm{SiO}_2$  sturcutre, and extracted experimental results for 50 nm (black stars), 75 nm (red stars), and 110 nm (green stars) nanocubes. Graphene nonlocal conductivity model was used, with Fermi-level of 0.47eV, lifetime of 10fs, and T=300K for room temperature, and the dashed orange curve represents the Fermi velocity.
\label{fig:figure2}}
\end{figure*}

We further examine the device response by studying the gate-dependent extinction spectra obtained for different nanocube sizes and concentrations. The FTIR extinction spectrum was measured for three different nanocube sizes, of $50$ $\mathrm{nm}$, $75$ $\mathrm{nm}$, and $110$ $\mathrm{nm}$, as presented in Fig.~\ref{fig:figure2}A, Fig.~\ref{fig:figure2}B, and Fig.~\ref{fig:figure2}C, respectively. For each nanocube size, two samples were fabricated and measured, one with a higher nanocube concentration (solid curves) and one with a lower nanocube concentration (dashed curves) \cite{Si2020	}. Similar to the behavior in Fig.~\ref{fig:figure1}C, for all nanocube sizes the resonances are seen to shift to higher frequencies with increasing graphene doping. An overall shift of the resonances to lower energies with increasing nanocube size can also be seen, which is attributed to the dispersive nature of AGPs (see discussion below). In addition, a strong hybridization with the $\mathrm{SiO}_2$ surface-phonon occurs when the resonance is close to the $\mathrm{SiO}_2$  phonons, due to the nanocube size and/or charge carrier density in graphene. This hybridization results in an enhancement of the peak around $1120$ $\mathrm{cm^{-1}}$ due to the strong oscillator strength of the $\mathrm{SiO}_2$  phonon. \\

The calculated and experimental dispersion relation, i.e. energy vs. momentum dependence, is compared in Fig.~\ref{fig:figure2}D, showing good agreement. For simplicity, we calculate the dispersion for the layered structure without geometrical features (colormap), i.e. Au/dielectric spacer/graphene/$\mathrm{SiO}_2$ , and the experimental momentum (colored markers) is calculated by $2\pi/\mathrm{L}$, where $\mathrm{L}$ is the nanocube length. It can be seen that the resonances for both 75 nm and 110 nm nanocubes (green and red markers, respectively) lie very close to the $\mathrm{SiO}_2$ phonons, thus exhibiting a more phonon-like nature, confirming the strong hybridization mentioned above. The resonances for 50 nm nanocubes (black markers), however, are farther away in frequency from the $\mathrm{SiO}_2$  phonons, thus displaying a more plasmon-like nature. This also explains the larger shift of the resonances with graphene doping that is obtained for the 50 nm nanocubes, as compared to the 75 nm and 110 nm nanocubes, stemming from the fact that AGPs are doping dependent while the $\mathrm{SiO}_2$  phonons are not. It is also seen that the calculated dispersion lies close to the Fermi velocity, $\mathrm{V}_\mathrm{F}=1\cdot10^{6}$ m/s (dashed orange line), which corresponds to the lowest velocity that AGPs can be slowed down to, and to the maximal wavelength confinement of $\lambda_0/300$. For the closest obtained experimental point (lower black marker), we calculate an AGP velocity of $\mathrm{\sim}1.42\cdot10^{6}$ m/s, denoting a very strong confinement of the optical field.  \\

The results obtained in Fig.~\ref{fig:figure2} further show that for different concentrations of the same nanocube size, one obtains the same spectral response but with different amplitude. On the other hand, different nanocube sizes generate an overall different spectral response. This implies that it is the single nanocube properties that determines the optical response of the device, while the amount (concentration) of nanocubes determines its amplitude. This is reinforced by the previous observation of the weak interaction between neighboring nanocubes. We can therefore conclude that a single nanocube/hBN/graphene structure actually acts as a single resonator. \\

We note that in these samples, the measured fill-factor, which is the percentage of area covered by the nanocubes, ranges between $3\%-13\%$ \cite{Si2020}. Yet, even with such low fill-factors the optical response is still comparable in magnitude to that obtained for grating couplers, with an equivalent fill-factor of $50\%-85\%$ \cite{Iranzo2018}. Furthermore, this robust scheme does not require any lithographic processes or patterning, and thus both preserve the graphene quality and removes any limitation on the device areas. To increase the number of devices per sample, we limited the graphene area to stripes of $200$ $\mathrm{\mu m}$ X $ 2 $ $\mathrm{mm}$, however, $\mathrm{cm}$-scale devices were also fabricated and measured \cite{Si2020}. \\

\begin{figure*}[p!] 
  \centering
  \includegraphics[scale=0.65]{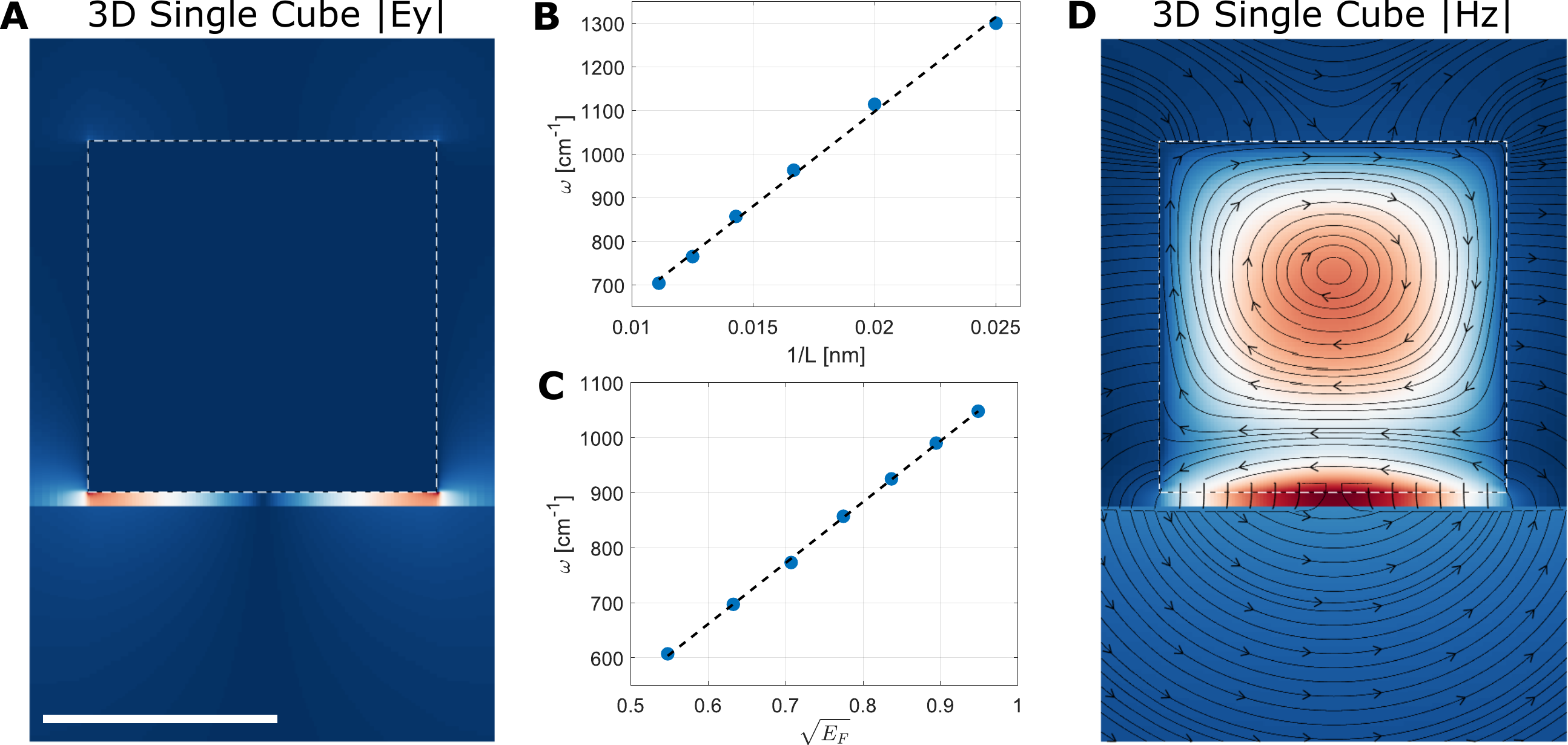} 
   \caption{\textbf{3D simulations and analysis of the generated electromagnetic fields of a single GPMR}. \textbf{A} The simulated $\mathrm{|E_y|}$ field distribution over the cross-section of a single 3D nanocube. The intense electric-field formed between the nancoube and the graphene is a signature of the generated AGPs (scale bar is $50$ $\mathrm{nm}$). \textbf{B} The simulated resonance frequency obtained for different nanocubes length $\mathrm{L}$ (between 40-90 nm), and \textbf{C} square-root of the Fermi-level, $\sqrt{\mathrm{E}_{\mathrm{F}}}$ (between 0.3-0.9 eV), respectively, showing the expected linear dispersion of AGPs. \textbf{D} The simulated magnetic field distribution, $\mathrm{|H_z|}$, superimposed with the electric field lines, showing the generation of a magnetic dipole resonance at the graphene/nanocube interface. Vacuum is used as the environment, and a 3 nm spacer thickness. 
\label{fig:figure3}}
\end{figure*}

To corroborate the single nature of the behavior, we now turn to the examination of the optical response of a single structure. We perform 3D simulations of a single nanocube close to a graphene sheet, which is illuminated by a far-field free-space beam \cite{Si2020}. The spatial distribution of the electric field $\mathrm{|E_y|}$, is presented in Fig.~\ref{fig:figure3}A, showing that the field is mainly confined between the graphene and the nanocube, as expected for AGP modes \cite{Iranzo2018,Lee2019,Alonso-Gonzalez2017}. In addition, Fig.~\ref{fig:figure3}B and Fig.~\ref{fig:figure3}C show the obtained simulation results of the resonance frequency for different nanocube dimensions and Fermi-levels, respectively. A linear dispersion relation is clearly seen in Fig.~\ref{fig:figure3}B, further corroborating the AGP nature of the resonance, together with a linear dependency on $\sqrt{\mathrm{E}_\mathrm{F}}$, which is characteristic to all plasmons in graphene. The field distribution obtained in Fig.~\ref{fig:figure3}A also exhibits the ability of the single structure to directly excite AGPs from the far-field. However, still remains the important question of what is the physical coupling mechanism, which enables the far-field excitation of such high momentum modes, with a completely random structure that is built from single cavities? \\


It can be argued that the average inter-nanocube spacing might lead to a certain periodicity in the system, which can contribute momentum similar to a grating coupler. However, we see no change in the spectral response of different concentrations (Fig.~\ref{fig:figure2}), which is contradictory to what has been observed for periodical structures with different periods \cite{Iranzo2018,Lee2019}. Owing to these observations, we can safely rule out the existence of any order-based momentum-matching condition. A further examination of the magnetic field distribution of the single structure reveals its coupling nature. Fig.~\ref{fig:figure3}D shows the spatial distribution of the magnetic field for a single nanocube close to a graphene sheet, superimposed with the electric field lines. The lines around the graphene/nanocube interface form a loop that is correlated with a strong magnetic field in its center, which has the shape of a magnetic dipole resonance. In the radio-frequency regime, this type of behavior occurs when a rectangular metallic patch is placed above a grounded conductive plane, and is known as the patch antenna. It supports Fabry-Perot-like resonances and can be described by a magnetic surface current. For conductive patch and ground plane, this electromagnetic response is not constrained to a specific spectral band. Indeed, metal nanocubes placed close to a gold surface, known as the nanocube-on-metal (NCoM) system, had been previously shown to act as magnetic resonators in the VIS spectrum \cite{Moreau2012}. \\


The optical response of the GPMR structure now becomes very clear. Graphene, being a semi-metal, can both act as a conductor and support AGPs when sufficiently doped. Thus, if a nanocube is placed in its vicinity, the illuminating far-field light can directly excite the GPMR patch antenna mode, which is associated with the excitation of an AGP between the nanocube and graphene, forming an optical cavity. The scalable nature of the patch antenna can also be observed in the GPMR spectral response for different nanocube sizes, where larger nanocube size corresponds to a larger effective wavelength of the mode (see Fig.~\ref{fig:figure2} and its discussion). We note that the remaining magnetic field within the top part of the nanocube in Fig.~\ref{fig:figure3}D is the part left unscreened by the graphene, owing to its small thickness and lower charge carrier density, compared to a semi-infinite metal surface.\\

\begin{figure*}[p!] 
  \centering
  \includegraphics[scale=0.65]{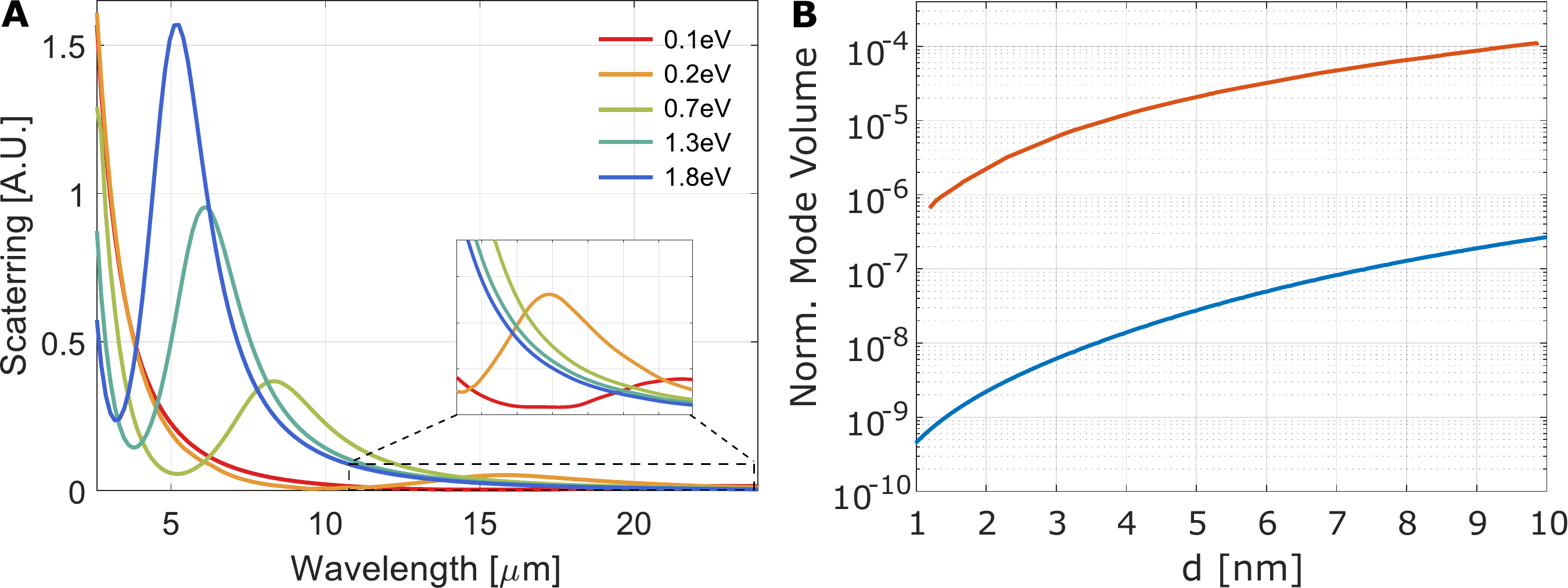} 
   \caption{\textbf{GPMR antenna response and mode-volume calculation}. \textbf{A} Simulated 3D scattering from a single GPMR antenna, for different graphene Fermi levels, showing that the scattering spectrum can be tuned from the far-infrared to almost the near-infrared. \textbf{B} The calculated normalized mode-volume of the GPMR cavity in the MIR (blue curve), for different nanocube-graphene spacing, d, and its comparison to the NCoM cavity in the visible spectrum (red curve), showing orders-of-magnitude smaller mode-volumes, with decreasing spacer thickness. In these simulations, the nanocube size is $75$ $\mathrm{nm}$ and hBN was used as the spacer. 
\label{fig:figure4}}
\end{figure*}

To corroborate the fact that the GPMR cavity is actually a graphene-based patch antenna in the MIR, we examine its scattering response. Owing to the similarity in the scattering response of patches and stripes \cite{Wu2011,Moreau2012}, for simplicity of simulations we show in Fig.~\ref{fig:figure4}A the simulated 2D scattering of a single GPMR structure, for several Fermi levels, ranging from $0.1\mathrm{eV}$ to the recently achieved $1.8\mathrm{eV}$\cite{Kanahashi2019}. Scattering resonances that are correlated with the AGP resonances can be observed and indeed validate the GPMR antenna nature. Strikingly, the response can be tuned from the far-infrared almost to the near-infrared spectrum, solely by changing the charge carrier density in the graphene. Such an ultra-broad spectral response is quite remarkable to obtain from a single antenna. Although a variety of graphene-based antennas have been proposed, mainly in the MIR and THz \cite{Jornet2013,Correas-Serrano2017}, they were found to be inefficient, compared to their metallic counterparts. Thus, the GPMR might provide a different solution for realization of nano-antennas at these spectral bands. \\

Finally, we study the mode-volumes achieved by the GPMR cavity using the quasi-normal mode (QNM) theory \cite{Si2020}. Fig.~\ref{fig:figure4}B shows the calculated normalized mode-volume, $\mathrm{V}_{\mathrm{GPMR}}/\lambda_{0}^{3}$ (with $\lambda_{0}$ being the free-space wavelength), of a single GPMR cavity with different hBN spacer thicknesses $\mathrm{d}$ (blue curve). Since similar plasmonic cavities in the visible	 spectrum have been shown to reach large confinement as well\cite{Akselrod2014,Chikkaraddy2016}, the GPMR cavity mode-volume is compared with the equivalent NCoM patch antenna in the visible spectrum, where graphene is replaced by a gold surface\cite{Moreau2012} (red curve). It can be seen that even though the GPMR wavelength is about an order of magnitude larger than the NCoM wavelength, it is able to achieve a normalized mode-volume which is $4$ orders of magnitude smaller, reaching $\sim4.7\cdot10^{-10}$. These remarkable values can be directly attributed to the unique properties of AGPs and thier ability of confining light to very small dimensions. \\

Although we can further thin down the hBN spacer until the monolayer case\cite{Iranzo2018}, we note that below $1$ $\mathrm{nm}$ thickness, strong nonlocal effects, in both metal and graphene, dominate the confinement and enhancement factors that can be achieved\cite{Ciraci2012,Iranzo2018}. These require a special analytical treatment that can not be introduced into our numerical simulations. As it has been previously shown that AGPs can confine light to a monolayer hBN spacer and without increasing the losses \cite{Iranzo2018}, thinning down the spacer should continue and improve the achievable mode-volume presented in Fig.~\ref{fig:figure4}b	.\\


In conclusion, we have demonstrated the realization of single AGP cavities in the MIR spectrum with record small mode-volumes, which are efficiently excited from the far-filed. Our approach provides a tunable platform for studying strong light-matter interactions in the MIR and THz spectra, such as vibrational strong-coupling \cite{Thomas2019,Autore2018}, and its manipulation of chemical processes\cite{Thomas2019}. Furthermore, it opens up new possibilities for the realization of efficient AGP-based devices in this long wavelength spectrum, such as photodetectors, biological and chemical sensing devices, and graphene-based tunable optical nano-antennas. \\

\section*{\textbf{Author Contribution}}
I.E. conceived the idea, performed simulations, experiments and analysis of the results. D.A., A.K., and V.P. fabricated devices and assisted in measurements. Z.H. and X.D. preformed the nanocubes deposition and its analysis, T.K. assisted in device fabrication, N.P., T.R and J.P.H assisted in numerical simulations, N.P., D.R.S. and F.K. supervised the project. All authors contributed to the writing of the manuscript. \\

\begin{acknowledgments}
{I.E. would like to thank Eduardo J. C. Dias for fruitful discussions and Dr. Fabien Vialla. D.R.S acknowledges the support of AFOSR(FA9550-12-1-0491, FA9550-18-1-0187) grants. N.M.R.P. acknowledges support from the European Commission through the project “Graphene-Driven Revolutions in ICT and Beyond”  (Ref. use CORE 3 reference, not CORE 2). N.M.R.P. and  T.G.R. acknowledge COMPETE 2020, PORTUGAL 2020, FEDER and the Portuguese Foundation for Science and Technology (FCT) through project POCI-01-0145-FEDER-028114. F.H.L.K. acknowledges financial support from the Government of Catalonia trough the SGR grant, and from the Spanish Ministry of Economy and Competitiveness, through the “Severo Ochoa” Programme for Centres of Excellence in R and D (SEV-2015-0522), support by Fundacio Cellex Barcelona, Generalitat de Catalunya through the CERCA program,  and the Mineco grants Ramón y Cajal (RYC-2012-12281,  Plan Nacional (FIS2013-47161-P and FIS2014-59639-JIN) and the Agency for Management of University and Research Grants (AGAUR) 2017 SGR 1656. Furthermore, the research leading to these results has received funding from the European Union Seventh Framework Programme under grant agreements no.785219 and no. 881603 Graphene Flagship. This work was supported by the ERC TOPONANOP under grant agreement no. 726001 and the MINECO Plan Nacional Grant 2D-NANOTOP under reference no. FIS2016-81044-P.
} \end{acknowledgments}

\bibliography{GPMR}

\end{document}